\documentclass[12pt]{iopart}
\usepackage{graphicx}

\begin{document}
\begin{flushright}
\today
\end{flushright}

\title{Growth and superconducting transition of Pr$_{1-x}$Ca$_x$Ba$_2$Cu$_3$O$_{7-\delta}$($x\approx$0.5) epitaxial thin films}
\author{J. Y. Xiang, Z. Y. Liu, J. Li, P. Wang, R. L. Wang, H. P.
Yang, Y. Z. Zhang, D. N. Zheng\footnote[3]{To whom correspondence
should be addressed (e-mail: Dzheng@ssc.iphy.ac.cn)}, H. H. Wen,
and Z. X. Zhao}
\address{ National Laboratory for Superconductivity,
Institute of Physics and Center for Condensed Matter Physics,
Chinese Academy of Sciences, Beijing 100080, People's Republic of
China. }

\begin{abstract}
Pr$_{1-x}$Ca$_x$Ba$_2$Cu$_3$O$_{7-\delta}$($x\approx$0.5) thin
films have been grown on SrTiO$_3$ and YSZ substrates by the
pulsed laser ablation. The substrate temperature dependence of
orientation and superconducting poperties were systematically
studied. Good quality \textit{c} and {a}-axis orientated films can
be obtained on SrTiO$_3$ via changing the substrate temperature
solely. On YSZ, films with good \textit{c}-axis orientation can be
grown, while it is hard to grow films with good \emph{a}-axis
orientation by changing substrate temperature alone. The highest
$T_{C0}$ is about 37K, which is found in the films grown on YSZ
with a good \emph{c}-axis orientation. For the films grown on STO,
however, the highest $T_{C0}$ is about 35.6K with a mixed
orientation of \emph{c}-axis and \emph{a}-axis. In most of the
superconducting films, the weak temperature dependence of the
normal state resistivity, as characterized by small
$R(290K)/R(50K)\leq2$ ratios, together with a weak localization
behavior just above $T_C$ could be attributed to the essential
scattering due to the localized electronic states. The
superconducting transitions in a field up to 10 T along
\textit{c}-axis have been measured on a \textit{c}-axis oriented
film grown on SrTiO$_3$. The zero-temperature in-plane upper
critical field $B^{ab}_{C2}$(0) is estimated from the resistivity
transition data.
\end{abstract}

 \submitto{\SUST}
\maketitle

\section{Introduction}
Among the high $T_C$ cuprates, compounds with $\rm
RBa_2Cu_3O_{7-\delta}$ (\textit{R}=rare-earth elements) structure,
the so-called 123 structure, have properties which are nearly
independent of R except for \textit{R=Pr}\cite{Hor,Fisk}. Although
$\rm PrBa_2Cu_3O_{7-\delta}$ (\textit{Pr-123}) can be formed, it
does not show superconductivity at low temperature and even shows
no metallic behavior\cite{soderholm}. There are also reports on
superconductivity in \textit{Pr-123}\cite{blackstead,zou},
however, the bulk superconductivity of \emph{Pr-123}, has to be
independently verified and also the structure of superconducting
\emph{Pr-123} has to be confirmed by rigorous crystallography.
Optical study has shown that $\rm PrBa_2Cu_3O_{7-\delta}$ is a
charge transfer type insulator with a charge-transfer gap
$\Delta_{CT}\sim$1.4 eV, and carriers in the chain are localized
at low temperatures and low frequencies~\cite{takenaka}. Due to
the depletion of mobile carriers, the in-plane electric transport
properties in \textit{ Pr-123} could be described in terms of
variable range hopping (VRH) in the insulating region
~\cite{Fisher}. When the high-temperature cuprates are doped with
$Pr$ the superconductivity is usually suppressed~\cite{Akhavan},
and often a metal-insulator transition occurs, accompanied by the
presence of complicated magnetic behavior due to the relatively
large magnetic moment of Pr ions~\cite{boothroyd}. The origin of
the suppression of superconductivity in HTSC by \textit{Pr} and
absence of superconductivity in \textit{Pr-123}, being of
fundamental interest, may also shed light on the mechanism of
high-temperature superconductivity, since any reasonable theory
should explain the effect of \textit{Pr} on superconductivity in
123 systems.

If the carriers depletion were indeed the main origin for the
absence of superconductivity in \textit{Pr-123}, a simple approach
could be to introduce carriers by chemical doping. A substitution
of divalent element such as Ca for Pr appears to be a natural
choice. Nevertheless no superconductivity was found in the samples
with fairly low Ca doping levels (less than 30\% at ambient
pressure\cite{hdyang,yfxiong}). Norton \textit{et al.}
successfully revived superconductivity in Pr-123 epitaxial thin
films by substituting 50\% of the tetravalent Pr by divalent
Ca~\cite{NortonA,NortonB}. And later, superconductivity was again
found in high-pressure synthesized Ca-doped Pr-123 bulk
samples~\cite{ZXZhao,ysyao}. However, the influence of crystal
structures on superconductivity in the thin films is still
somewhat ambiguous. For example, it is necessary to study the
superconductivity in films with different orientations
systematically. And also it is important to make a comparative
study of the transport properties of thin films and that of
ceramic bulk samples. The transport properties of Ca-doped Pr-123
bulk samples sintered at ambient pressure have been studied by
Luszczek~\cite{MLuK}. However, researches are harassed by the
limited Ca substitution levels and no superconductivity was
observed. Whereas, for the samples synthesized at high pressures,
impurity phases seems hard to be avoided, although the Ca content
is highly improved~\cite{kqli}. In this article, we report a
systematical investigation on the substrate temperature dependence
of film orientation, the superconducting transition temperature
and the residual-resistivity ratio for $\rm
Pr_{0.5}Ca_{0.5}Ba_2Cu_3O_{7-\delta}$ thin films grown on the
substrates of SrTiO$_3$ and Yttrium-stabilized ZrO$_2$. The
superconducting transition data of $\rm
Pr_{0.5}Ca_{0.5}Ba_2Cu_3O_{7-\delta}$ epitaxial thin films
measured in a magnetic field up to 10 T, applied parallel to
\textit{c}-axis of the films, are analyzed.

\section{Film growth and characterization}
A $\rm Pr_{0.5}Ca_{0.5}Ba_2Cu_3O_{7-\delta}$ ceramic target was
prepared using a conventional solid state reaction method similar
to that used to produce superconducting R-123 pellets.  The
stoichiometric quantities of high-purity dry $\rm Pr_6O_{11}$,
$\rm CaCO_3$, $\rm BaCO_3$, and CuO powders were ground, mixed,
fired, pelletized and sintered for several times. The sintering
temperature was 930 $^{\circ}$C. The whole process was carried out
in air under an ambient pressure. Powder X-ray diffraction
indicated a predominant R-123 phase while a small amount of $\rm
BaCuO_2$ could also be detected. The $\rm
Pr_{0.5}Ca_{0.5}Ba_2Cu_3O_{7-\delta}$ epitaxial thin films were
grown using pulsed-laser ablation mainly on the substrates of
SrTiO$_{3}$ (STO) and Yttrium-stabilized ZrO$_2$ (YSZ). Some
LaAlO$_3$ (LAO) and (LaAlO$_3$)$_{0.3}$ -(Sr$_2$AlTaO$_8$)$_{0.7}$
(LSAT) substrates were also used. A silicon heater was utilized in
the film deposition, and the temperature was monitored by an
infrared thermodetector with a calibrated temperature range from
300 $^{\circ}$C to 1200 $^{\circ}$C. A self-made NiCr-NiAl
thermocouple was placed at the back of the heater as a temperature
reference. A LPX 300 KrF ($\lambda$ =248 nm) excimer laser (Lambda
Phyisk) was adopted for the film growth. The repetition rate was 5
Hz and the fluence was about 2 J/cm$^2$. The target-substrate
distance was kept at 70 mm. The films were deposited in an oxygen
partial pressure of 40 Pa, as frequently seen in the growth of
YBCO. After deposition, a one-minute holding before annealing was
used for the purpose of stress relaxation. The annealing process
was carried out in 1 atm oxygen at a cooling rate around 15 $\rm
^o$C/min for about 10 minutes (The superconductivity of the film
was found hardly dependent on the annealing duration at this
circumstance). The films were then cooled down naturally to room
temperature. The superconducting critical temperature of the films
was measured using a standard four-probe technique, with the
electrodes mounted by silver paste on the films directly.

\begin{figure}
\includegraphics[height=\columnwidth,angle=-90]{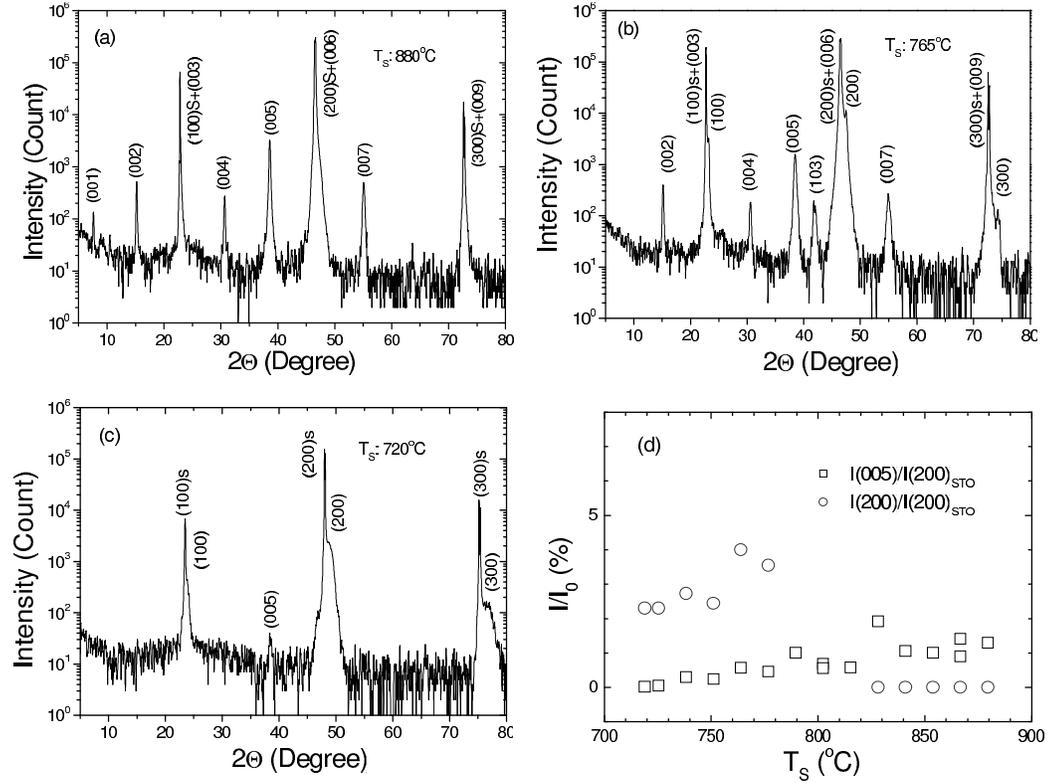}
\caption{\label{fig1}Influence of the substrate temperature $T_S$
on the orientation of $\rm
Pr_{0.5}Ca_{0.5}Ba_{2}Cu_{3}O_{7-\delta}$ thin films grown on STO.
Typical XRD patterns for films grown at $T_S$ of (a) 880 $\rm
^o$C, (b) 765 $\rm ^o$C, and (c) 720 $\rm ^o$C are shown. Peaks of
the substrate are marked with a postfix "s". Shown in (d) is the
$T_S$ dependence of the ratio I(005)/I(200)$_S$ and
I(200)/I(200)$_S$, where I(\textit{hkl}) is the diffraction
intensity of (\textit{hkl}) peak for the film or the substrate.}
\end{figure}

\begin{figure}
\includegraphics[width=0.88\columnwidth]{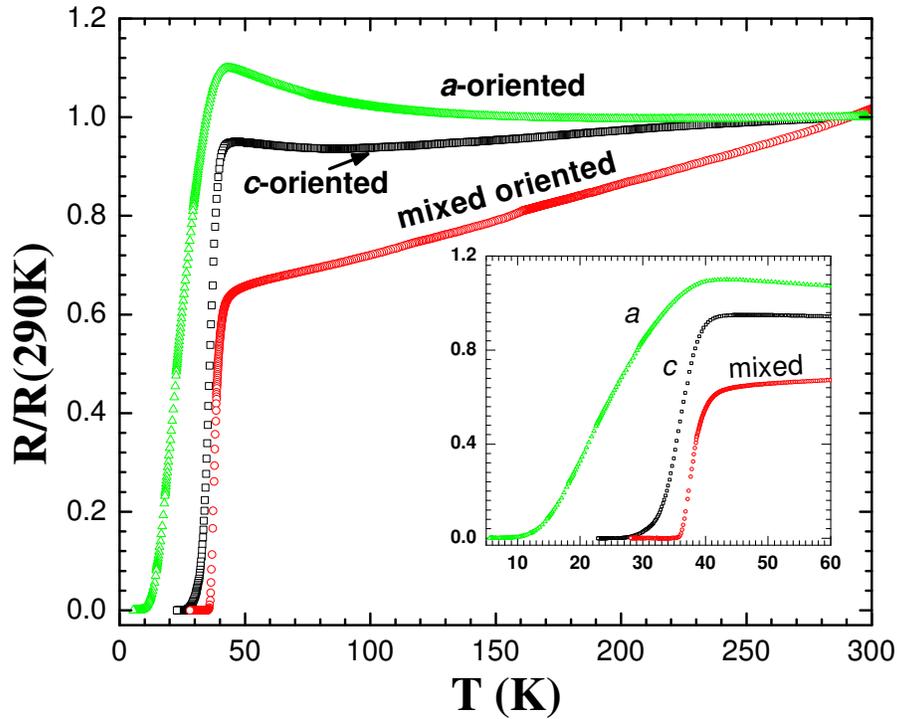}
\caption{Typical resistance transition of different oriented films
grown on STO. The resistance of these films are normalized to the
value at 290 K, \textit{i.e.}, R(290K). Here, labels with
'\textit{c}-oriented', 'mixed oriented' and '\textit{a}-oriented'
correspond to a substrate temperature of 880 $\rm ^o$C, 790 $\rm
^o$C and 720 $\rm ^o$C, respectively. Inset is the low temperature
part of the transition of these films.} \label{fig2}
\end{figure}

\begin{figure}
\includegraphics[width=0.88\columnwidth]{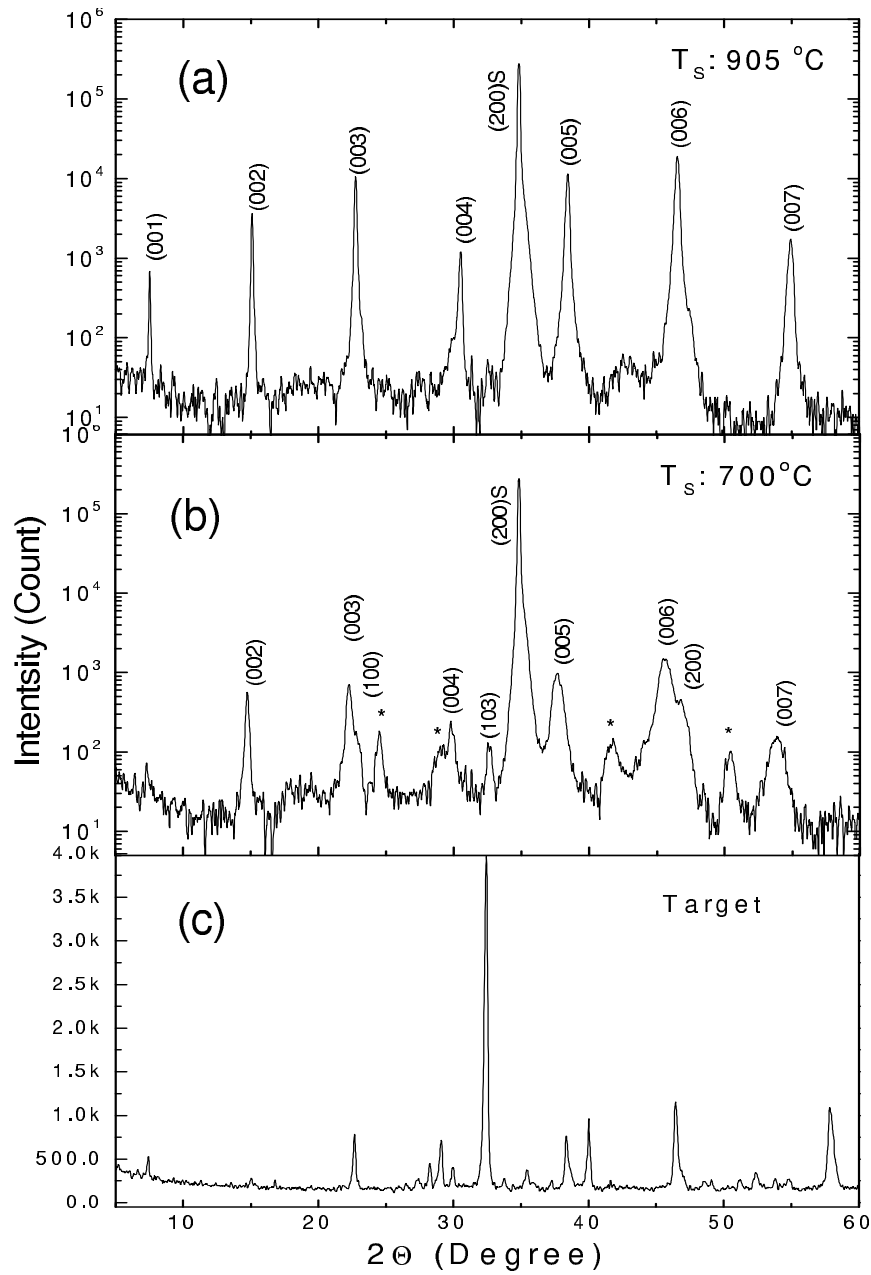}
\caption{Substrate temperature $T_S$ dependence of the crystal
structure for films grown on YSZ. (a) and (b) are the XRD patterns
correspond to films grown at $T_S$ of 905 $\rm ^o$C and 700 $\rm
^o$C, respectively. Star in (b) denotes the mis-orientation or/and
the secondary phase. As a comparison, the XRD pattern of the
target is shown in (c).}\label{fig3}
\end{figure}

Inductively coupled plasma (ICP) results show a good consistency
of the film stoichiometry with that of the target, as expected
from the pulsed-laser ablation method. The orientation of the
films was characterized by X-ray diffraction (XRD). On STO, films
with \textit{c}-axis lying in the film plane (\textit{a}-axis
oriented) can be obtained by decreasing the substrate temperature
$T_S$ for about 160 $\rm ^o$C, as compared with that for films
with \textit{c}-axis perpendicular to the film surface
(\textit{c}-axis oriented). Typical XRD patterns for films with
\textit{c} and \textit{a}-axis orientation, grown at a temperature
of 880 $\rm ^o$C and 720 $\rm ^o$C, are shown in Fig.~\ref{fig1}a
and Fig.~\ref{fig1}c, respectively. Films grown at an intermediate
$T_S$ of $\sim740~\rm ^o$C possess a structure with mixed
\textit{c} and \textit{a}-axis orientated phases, as shown in
Fig.~\ref{fig1}b. A weak (103) peak, which is the strongest line
of the target spectrum, can sometimes be seen as $T_S$ was further
reduced. The surface profile of the films were obtained by atomic
force microscopy (AFM). The mean roughness is around 4 nm and 20
nm in a 10$\times10~\mu m^2$ area for the \textit{a} and
\textit{c}-axis oriented films, respectively. As a comparison, the
intensity ratios of (005) and (200) of the film to (200) of the
substrate, varying with $T_S$, are also shown in Fig.~\ref{fig1}d.
Typical resistance transitions for the films with different
orientation are shown in Fig.\ref{fig2}. Here, the resistance of
the films scaled by the respective value at 290 K, R(290K). Label
'\textit{c}-oriented', 'mixed oriented' and '\textit{a}-oriented'
of the curves corresponds to a substrate temperature of 880 $\rm
^o$C, 790 $\rm ^o$C and 720 $\rm ^o$C, respectively. The low
temperature part of transition of the curves are also zoomed in
the inset of Fig.\ref{fig2} for a clearer view. It is evident that
the film structure has a substaintial effect on both
superconducting and normal state properties of the film. On YSZ
substrates, good quality \textit{c}-axis oriented films can also
be obtained, as shown in Fig.~\ref{fig3}. The substrate
temperature in such a growth process was around 900 $\rm ^o$C,
slightly higher than that in the case of STO. It can be attributed
to the 45 $\rm ^o$ rotation of the $R-123$ unit cells on YSZ.
However, it was found difficult to grow \textit{a}-axis films on
YSZ solely by varying $T_S$. When $T_S$ was reduced lower than
that for \textit{a}-axis oriented films grown on STO, more and
more characterized peaks of the target appeared. A typical XRD
pattern for the films grown at about 700 $\rm ^o$C is shown in
Fig.~\ref{fig3}b. We can see that peaks other than the
(00\textit{l}) and (\textit{l}00) are shown. For substrates such
as LAO and LSAT, good quality \textit{c}-axis oriented films can
also be obtained.

\begin{figure}\begin{center}
\includegraphics[width=0.88\columnwidth]{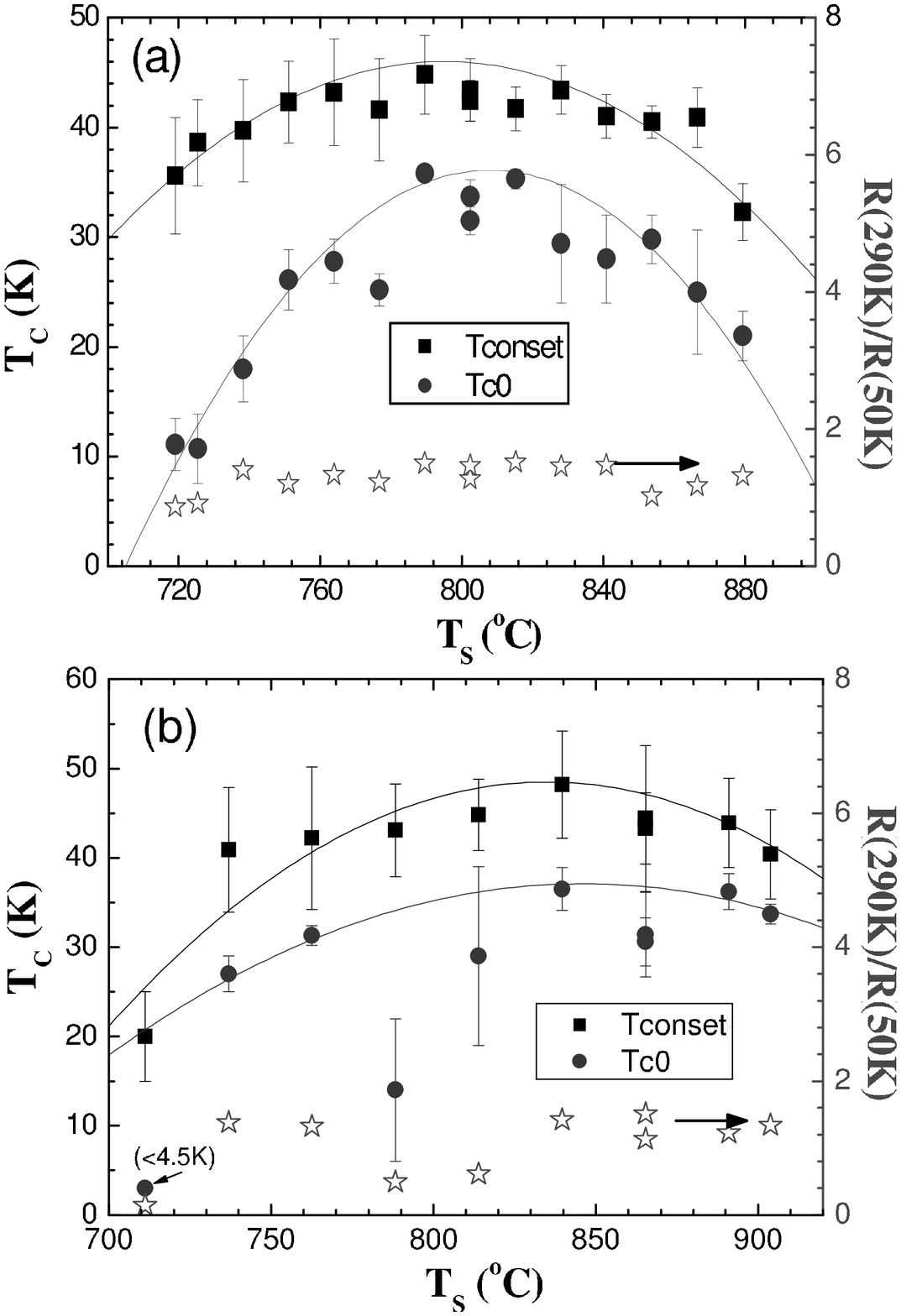}
\caption{$T_S$ dependence of the superconducting transition
temperature $T_{C0}$ and $T_{Conset}$ for the $\rm
Pr_{0.5}Ca_{0.5}Ba_{2}Cu_{3}O_{7-\delta}$ films grown on (a) STO
and (b) YSZ. Definitions of $T_{C0}$ and $T_{Conset}$ are detailed
in the text. The solid lines in both graphs are just guides to the
eyes. The hollow stars represent the ratio R(290K)/R(50K) in both
graphs as directed by the arrow. Thickness of all the films shown
here are around 800\AA~for STO and 1200\AA~for YSZ.}\label{fig4}
\end{center}\end{figure}

It is found that the critical temperature changes with the
substrate temperature. Fig.~\ref{fig4} shows the substrate
temperature $T_S$ dependence of the critical temperature for films
grown on STO. The error bar of $T_{Conset}$ and $T_{C0}$
corresponds to the temperature scope from 100\% to 90\% of
$\rho_n$, and 10\% to 0 (within the resolution of the instruments
used) of $\rho_n$, respectively. Here $\rho_n$ is the normal state
resistivity, defined as the value where the sample resistivity
deviates from the linear temperature dependence near the
superconducting transition. The relationship of $T_C$ and $T_S$
for the films on YSZ is also shown. A higher $T_S$ is necessary to
grow epitaxial thin films on YSZ than on STO, due to the larger
lattice mismatch of the former. Accordingly, $T_C$ for the films
on YSZ is higher, which could also be due to a larger strain at
the interface. The highest $T_{C0}$ is about 35.6 K and 37 K for
the films grown on STO and on YSZ, respectively. It is worthy to
notice that on STO, films with higher $T_{C0}$ are those with a
structure of mixed \textit{c}-axis and \textit{a}-axis oriented
phases, as was reported earlier by Norton \textit{et
al.}~\cite{NortonA}. For the films grown on YSZ, the variation
tendency of the measurement data seem to suggest a better
superconductivity for the \textit{c}-axis oriented films than
those with \textit{a}-axis orientation.

An interesting thing of the films is their metastable properties. As the solubility of Ca in the bulk materials
sintered at ambient pressure is less than 30\%, metastable properties of the target with higher Ca doping levels
could be expected, as having been pointed out by Norton \textit{et al.}~\cite{NortonB}. Sintering the bulk
materials under high pressure can raise the Ca substitution level to 70\%. Superconductivity exhibits in bulks
with Ca doping level from 30\% to 70\%, as mentioned previously in the literature~\cite{kqli}. However, the
superconducting bulk materials became insulating after fired in a furnace in air at a temperature higher than 200
$\rm ^o$C for half an hour~\cite{che}. Comparatively, by real-time monitoring using the reflection high energy
electron diffraction (RHEED) system, we found that it took less than 1 minute for the film diffraction pattern to
vanish, when the film was placed in a background pressure 10$^{-4}$ Pa at a temperature higher than 200 $\rm ^o$C.
There is nothing but some spots remained on the substrate after such a heating process, which can be observed by
the naked eyes. The spots are found insulating (The resistance read from a multimeter is in the magnitude of
M$\Omega$).

\section{Superconducting transition in a magnetic field}

The in-plane resistance measurement was carried out using the
standard four-probe method on one of the \textit{c}-axis oriented
films grown on STO. Thickness of the film is around 800 \AA. The
current density used was 700 A/cm$^2$. Low ohmic contacts were
made to the sample via coating the electrodes with Pt, and using
silver paste to attach gold lines to the electrodes firmly.

\begin{figure}
\includegraphics[width=0.88\columnwidth]{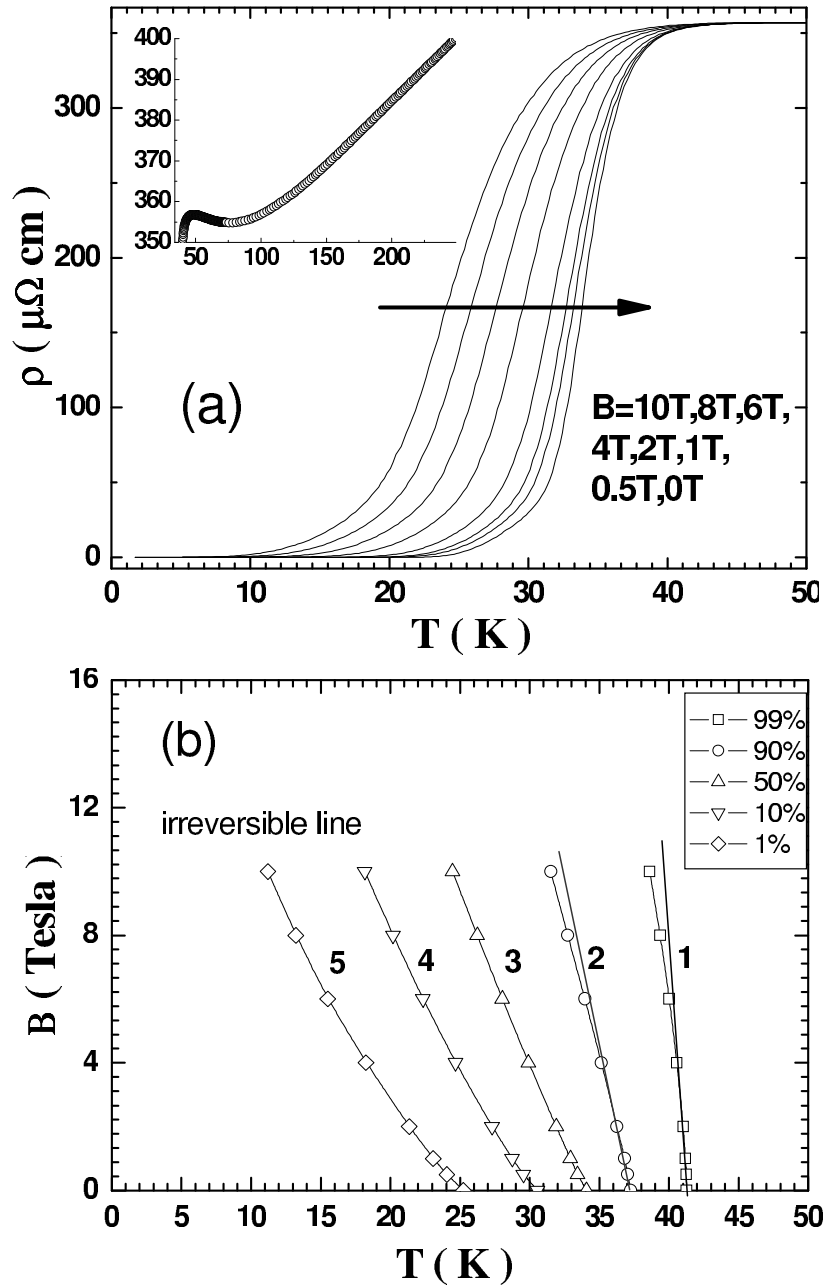}
\caption{(a) Temperature dependence of the film resistivity in
various magnetic fields H$\parallel$\textit{c}. A part of the R-T
curve measured at zero magnetic field is shown in the inset. (b)
H-T phase diagram for the epitaxial film at different criterions
1\%, 10\%, 50\%, 90\%, and 99\% of $\rho_n$, corresponding to line
1, 2, 3, 4, 5 respectively. $B_{C2}$(0) is estimated using the WHH
model to be $\simeq 105.5$ T and $\simeq47.1$ T for line 1 and
line 2, respectively. Film thickness here is about 800~\AA}
\label{fig5}
\end{figure}

A part of the $R-T$ curve measured at zero field is shown in the
inset of Fig.~\ref{fig5}a. The critical temperature $T_{Conset}$
is about 47 K with a transition width about 6K (from 90\% to 10\%
of $\rho_{n}$), which is close to other
reports~\cite{NortonA,NortonB,ZHWang}. However, the film has a
smaller resistivity just above $T_C$,
$\rho(T|_{T_C})=357~\mu\Omega $cm. From the graph we can see that
the transport behavior is metallic at high temperatures, but at
temperatures just above $T_{c}$, the resistance has a very small
enhancement. This should not be a contribution from the resistance
along \textit{c}-axis direction, since in the film \textit{a}-axis
oriented phase can be fully ruled out according to the XRD data.
We would rather attribute the effective enhancement to some kind
of weak localizations, although at present we do not have enough
evidence to determine what kind of localization it is. We propose
that in the film there are some localized charge carriers, so the
small enhancement is the combined result of the free carriers
itinerating and the localized carriers variable range hopping.

Superconducting transitions of the film in a field of 0.5, 1, 2,
4, 6, 8, and 10 T are presented in Fig.~\ref{fig5}a. For a more
clear inspection, we plot the H-T phase diagram of the film in
Fig.~\ref{fig5}b using the criterion 1\%, 10\%, 50\%, 90\%, and
99\% of $\rho_n$ respectively. Line $1$ is very close to the
irreversible line and line $5$ is very close to the upper critical
field.

\section{Discussions}

A common feature in the film transport behavior is that the
in-plane resistivity ratio R(290K)/R(50K) is less than 2 (more
precisely is 1.6) for the films grown on both STO and YSZ, as
indicated by the hollow stars in Fig.~\ref{fig4}. Before the
superconducting transition, resistivity of these films raises
slightly, suggesting a weak carrier localization. The small
resistivity slope and the localization-like behavior near the
superconducting transition temperature, which are often observed
in 50\% Pr doped YBCO, are almost independent of the magnetic
field strength, as being noticed from the resistivity data at B
$\leq$ 10 T. The weak localization behavior could be attributed
mainly to the low carrier density, as having been well shown in
the underdoped YBCO and $\rm La_{1-x}Sr_{x}CuO_4$ systems. In
fact, Hall measurements carried out in 50\% Ca-doped samples shows
an in-plane carrier density lower than that in the optimum-doped
YBCO~\cite{ZHWang,jyxiang}. The weak temperature dependence of
normal resistivity indicates an important scattering to the
electron from the localized states in the present system. The
disorder of Ba and Pr elements could contribute to such localized
states, which is possible in the present film grown in an oxygen
partial pressure of 40 Pa. The Pr ions may substitute for Ba in
the lattice, similar to the Nd and Ba disorder observed in $\rm
Nd_{1+x}Ba_{2-x}Cu_3O_7$ films grown at oxygen partial pressures
higher than 1 bar~\cite{hwu}.

So far, it is not well understood why films with a mixed
orientation has a better superconductivity than the purely
oriented (\textit{c} or \textit{a}) films grown on STO. We may
attribute this to an improved oxygen intake process for the film
of a mixed orientations. In other words, the mobile carrier
concentration may be higher in the mixed orientation films than
the purely oriented ones, as illustrated by the change of the
normal state resistivity slope in Fig.\ref{fig2}. Another feature
we have noticed is that the small enhancement of the resistivity
near $T_C$ is weakened in the mixed orientation films, and shows a
nearly linear temperature dependence of the resistivity as
illustrated by the 'mixed oriented' line in Fig.\ref{fig2}. And
thus seems to suggest an occurrence of structural defects
dependent delocalization of the carrier.

We have made an estimate about the upper critical field $B_{C2}$
from the field-dependent superconducting transition measurements.
By using the criterion of the $99\%$ and 90\% of the normal-state
resistivity $\rho_{n}$ for the determination of $B_{C2}$, a linear
temperature dependence is obtained near $T_{Conset}$ with a slope
$dB_{C2}/dT$ estimated to be around 3.5 T/K and 1.73 T/K for line
1 and line 2 respectively, as seen in Fig.~\ref{fig5}b. Using the
Werthamer, Helfand, and Holenberg (WHH)~\cite{WHH} extrapolation
to low temperatures with
$B_{C2}(0)=0.73(-dB_{C2}/dT\vert_{T=T_C})T_{C}$, the
zero-temperature in-plane upper critical field is estimated to be
$B^{ab}_{C2}(0)\simeq105.5~ T$ for line 1 and
$B^{ab}_{C2}(0)\simeq47.1~ T$ for line 2.

\section{Summary}
We have grown $\rm Pr_{0.5}Ca_{0.5}Ba_2Cu_3O_{7-\delta}$ epitaxial
thin films on different substrates using the pulsed laser ablation
technique. Good quality \textit{a} and \textit{c}-axis oriented
films can both be obtained on STO via varying the substrate
temperature. On YSZ, films with good \textit{c}-axis orientation
have also been grown. The highest $T_{C0}$ is around 35.6 K and 37
K for films grown on STO and YSZ substrates, respectively. These
are consistent with the results of superconductivity in Ca doped
\emph{Pr-123} as reported by Norton \emph{et al.} previously. The
films show a metastable thermostability. The very small
R(290K)/R(50K) ratio together with the localization and a large
normal state resistivity of the films is perhaps caused by the
essential scattering due to the localized electronic states. The
superconducting transitions in a magnetic field up to $10$ Tesla
for one film grown on STO have been measured. The zero-temperature
in-plane upper critical field $B^{ab}_{C2}$(0) is also estimated
from the resistance data.

\ack{}

We are grateful to G.C. Che for the fruitful discussions. We would
also give thanks to S.L. Jia for help in transport property
measurements. This work is supported by the National Natural
Science Foundation of China (10174093, 10174091, 10221002), the
Ministry of Science and Technology of China (NKBRSF-G1999064604)
and Chinese Academy of Sciences.

\section*{References}

\end{document}